\documentclass[notitlepage, final, 10pt, pra, aps, twocolumn, floatfix, showpacs]{revtex4-1}
\pdfoutput=1

\usepackage[english]{babel}
\usepackage[protrusion=true, expansion=true]{microtype}     
\usepackage{lmodern}  
\usepackage{graphicx}
\usepackage{amsmath, amssymb, amsfonts,bm}
\usepackage{latexsym}
\usepackage{bbm} 
\usepackage{mathtools}
\usepackage{placeins}
\usepackage{color}       
\usepackage{braket} 
\definecolor{navyblue}{rgb}{0.0, 0.0, 0.5}
\usepackage[colorlinks=true, breaklinks=true, linkcolor=navyblue, urlcolor=navyblue, citecolor=navyblue]{hyperref}

\newcommand{\D}[1]{\text{d}#1}
\newcommand{\ketbra}[2]{\ket{#1}\bra{#2}}
\newcommand{\expe}[1]{\Braket{#1}} 
\newcommand{\expes}[1]{\braket{#1}}
\newcommand{\Tr}[1]{\textrm{Tr}\left( #1\right)}
\newcommand{\ii}[0]{\mathrm{i}}
\newcommand{\ee}[0]{\mathrm{e}}
\newcommand{\oa}[0]{\hat{a}}   
\newcommand{\orh}[0]{\hat{\rho}}

\begin{document}
\title{Theory of remote entanglement via quantum-limited phase-preserving amplification}
\author{Matti Silveri} \author{Evan Zalys-Geller} \author{Michael Hatridge} \author{Zaki Leghtas}  \author{Michel~H.~Devoret} \author{S.~M.~Girvin}
\affiliation{Departments of Physics and Applied Physics, Yale University, New Haven, Connecticut 06520, USA}


\date{\today}

\begin{abstract}
We show that a quantum-limited phase-preserving amplifier can act as a which-path information eraser when followed by heterodyne detection. 
This `beam splitter with gain' implements a continuous joint measurement on the signal sources. 
As an application, we propose heralded concurrent remote entanglement generation between two qubits coupled dispersively to separate cavities. 
Dissimilar qubit-cavity pairs can be made indistinguishable by simple engineering of the cavity driving fields providing further experimental flexibility and the prospect for scalability. 
Additionally, we find an analytic solution for the stochastic master equation\----a~quantum filter\----yielding a thorough physical understanding of the nonlinear measurement process leading to an entangled state of the qubits. 
We determine the concurrence of the entangled states and analyze its dependence on losses and measurement inefficiencies.
\end{abstract}

\maketitle

\section{Introduction}
Spatially separated objects can be entangled by the measurement backaction~\cite{Hatridge13, Blais_joint, Roch14} of a joint measurement. 
In the process of concurrent remote entanglement generation, two remote stationary qubits are first entangled with separate flying qubits and which-path information is erased from them by interference effects~\cite{Zeilinger98, Zukowski93}. 
A subsequent measurement of the flying qubits can then implement a joint measurement with a backaction that projects the stationary sources to an entangled state, demonstrated in various atomic and solid state systems~\cite{Polzik01, Chou05, Moehring07, Ritter12, Hofmann12, Bernien13, Narla16} through coincidence detection of photons~\cite{Kok05}. 
In this concurrent scheme, entanglement generation occurs purely by measurement backaction; the entangled qubits exchange no information, not even unidirectionally. 
There need be no causal connection between the qubits. 
This represents an important conceptual difference from consecutive remote entangling configurations~\cite{Kerckhoff09, Roch14} with unidirectional exchange of information due to both qubits seeing the same photon field.

Generating entanglement is a necessity for quantum communication, cryptography and computation~\cite{Kimble08, Briegel01, Devoret13}. 
The concurrent configuration promotes scalability and modularity of a quantum network, allowing entangling operations between arbitrary nodes through routing independently generated parallel signals to a quantum eraser. 
Compared to consecutive configurations, high entanglement fidelity is harder to achieve since the single-qubit information is more exposed to losses. 
However, the concurrent method provides better on-off ratio for the effective entanglement since, with use of directional elements, no parasitic signal could in principle propagate from one qubit to the other. 

We show that a quantum-limited phase-preserving amplifier can be used as a quantum eraser for the which-path information for concurrent microwave signals (Sec.~\ref{sec:eraser}). 
When followed by detection of both quadratures of the amplified output signal, this novel `beam splitter with gain' can implement continuous joint measurement on the remote signal sources. 
The quantum eraser configuration is general for systems involving continuous variables~\cite{Braunstein05, Gaussian12}.
For concreteness, in Sec.~\ref{sec:cqed-gen} we propose and analyze remote entangling for superconducting qubits coupled to traveling continuous microwave signals (see also Ref.~\cite{RoyDevoret15}). Analogously to the spatiotemporal mode shapes, the single-qubit information from dissimilar cavity-qubit pairs (unequal dispersive coupling or decay rates) is carried by the unequal temporal measurement amplitudes. However, because the measurement amplitudes depend on the cavity dynamics, they can be made indistinguishable through simple engineering of the cavity driving amplitudes, reinforcing the scalability and experimental flexibility of the entangling scheme. In Sec.~\ref{sec:qfilter} we derive an analytic solution for the qubits' stochastic measurement dynamics and analyze the fidelity of the resulting entanglement with realistic estimates before concluding in Sec.~\ref{sec:conc}. 

\section{Phase-preserving amplifier as a~which-path information eraser} \label{sec:eraser}
Quantum-limited phase-preserving amplification can be implemented through non-degenerate parametric amplification. In this process two distinct incoming modes, denoted here as the signal and the idler at the frequencies $\omega_{\rm si}$ and $\omega_{\rm id}$, are coupled to a strong pump mode by a nonlinear three-wave mixing element yielding amplified outgoing modes. We will first summarize the derivation of the amplifier input-output relations~\cite{Clerk} before analyzing the erasure of the which-path information. 

\subsection{Input-output relations}
\begin{figure}
\centering
\includegraphics[width=1.0\linewidth]{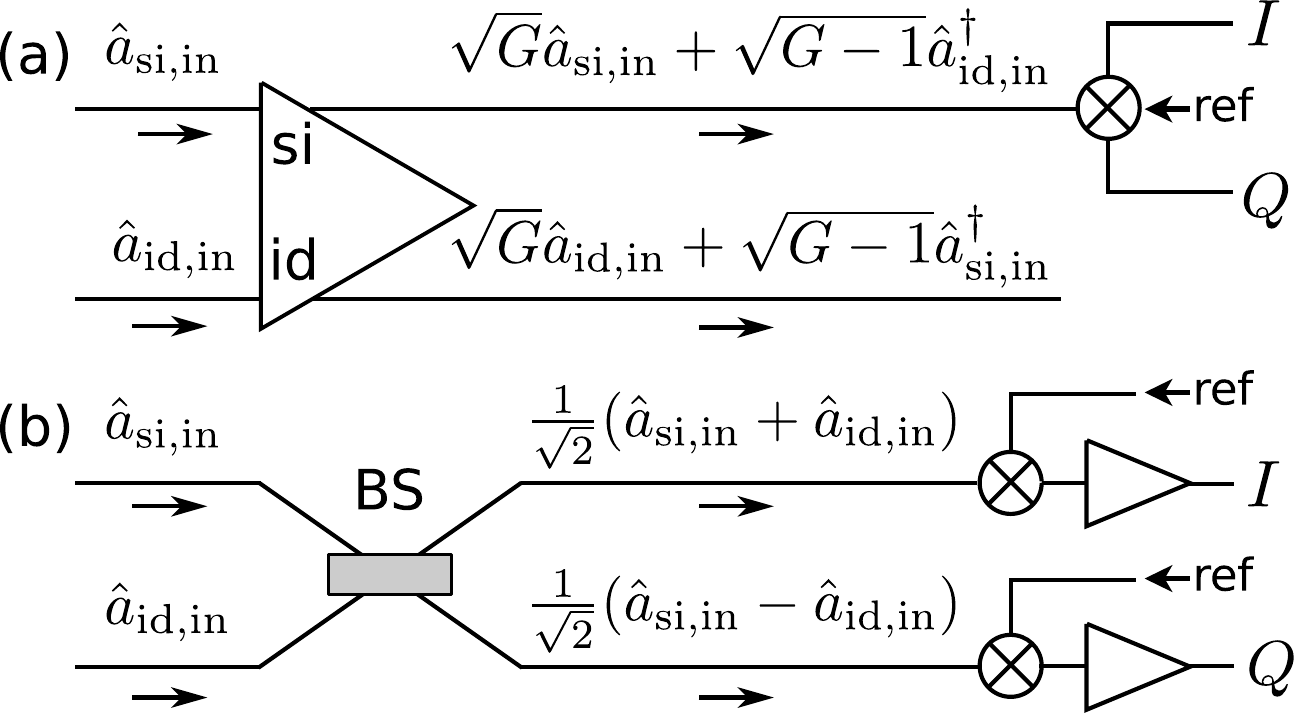}
\caption{\label{JPC_to_BS}(a)~Schematic of a quantum-limited phase-preserving amplifier~(large triangle) and subsequent heterodyne detection of both quadratures; (b)~An equivalent, high-gain representation using a $50$-$50$ beam-splitter (BS) and two phase-sensitive amplifiers~(triangles).
}
\end{figure}

We consider now a device with two ports and for each port we separate the incoming and outgoing modes. The device is operated in reflection but in the visualizations (see Figs.~\ref{JPC_to_BS}-\ref{cQED_schema}) we draw it in transmission for conceptual simplicity. 
For a reflective device, the signal input $\oa_{\rm si, in}(t)$ and output $\oa_{\rm si, out}(t)$ are related to the internal mode of the device $\oa_{\rm si}(t)$ through the coupling strength~$\kappa_{\rm si}$~\cite{GardinerCollet85}:
\begin{equation}
 \oa_{\rm si,out}(t)-\oa_{\rm si,in}(t)=\sqrt{\kappa_{\rm si}}\oa_{\rm si}(t). \label{inout}
\end{equation}
Because the internal mode is coupled to the external modes, it becomes damped at the rate $\kappa_{\rm si}$ and driven with $\sqrt{\kappa_{\rm si}}\oa_{\rm si, in}(t)$:
\begin{equation}
  \dot{\oa}_{\rm si}(t)=-\frac{\ii}{\hbar}[\oa_{\rm si}(t), \hat{H}_{\rm a}]-\frac{\kappa_{\rm si}}{2}\oa_{\rm si}(t)-\sqrt{\kappa_{\rm si}}\oa_{\rm si,in}(t). \label{internal}
\end{equation}
Similar equations hold for the idler mode $\hat{a}_{\rm id}(t)$. To solve for the outputs as a function of the inputs in Eqs.~\eqref{inout}-\eqref{internal}, we need to know the internal dynamics. For that purpose and for concreteness, we take the quantum-limited phase-preserving amplifier to be realized with a Josephson parametric converter amplifier (JPC)~\cite{Bergeal10, Bergeal10a, Abdo11}, whose Hamiltonian is a three-wave mixer, 
\begin{equation}
  \hat{H}_{\rm a}/\hbar=\sum_{k={\rm p, id, si}} \omega_k \oa^\dagger_k \oa_k-\ii\tilde{\lambda} \oa_{\rm si}^\dagger\oa_{\rm id}^\dagger\oa_{\rm p}+\ii\tilde{\lambda}^\ast \oa_{\rm si}\oa_{\rm id}\oa_{\rm p}^\dagger.
\end{equation}
To operate the device as an amplifier, the pump mode $\oa_{\rm p}$ is driven strongly at the frequency $\Omega=\omega_{\rm si}+\omega_{\rm id}$ such that it reaches a steady state $\oa_{\rm p}(t)=\alpha_{\rm p}\ee^{\ii\Omega t}+\oa'_{\rm p}(t)$. The pump mode provides the energy for the amplification. By ignoring the remaining quantum fluctuations in $\oa_{\rm p}(t)$ and going into the frame rotating at the eigenfrequencies of the signal and the idler modes, the resulting dynamics is set by the Hamiltonian
\begin{equation}
  \hat{H}'_{\rm a}/\hbar=-\ii \lambda \oa_{\rm si}^\dagger\oa_{\rm id}^\dagger+\ii \lambda^\ast \oa_{\rm si}\oa_{\rm id}, \label{H.lambda}
\end{equation}
where $\lambda=|\lambda|\ee^{\ii\varphi}=\alpha_{\rm p} \widetilde{\lambda}$ is the effective amplification strength. Using this Hamiltonian in Eq~\eqref{internal}, the input-output relations~\eqref{inout} can be expressed in the frequency domain as (see Ref.~\cite{Clerk} for details):
\begin{subequations}
\label{ainout_w}
\begin{align}
  \oa_{\rm si, out}(\omega)&=g(\omega)\oa_{\rm si,in}(\omega)+\ee^{\ii \varphi}\sqrt{|g(\omega)|^2-1}\oa^\dagger_{\rm id,in}(\omega),\\
  \oa_{\rm id, out}(\omega)&=g(\omega)\oa_{\rm id,in}(\omega)+\ee^{\ii \varphi}\sqrt{|g(\omega)|^2-1}\oa^\dagger_{\rm si,in}(\omega).
\end{align}
\end{subequations}
The gain factor $g(\omega)$ and the amplification bandwidth $D$ are
\begin{subequations}
\label{eq:gain_D}
\begin{align}
  g(\omega)&=\frac{\sqrt{G}-\ii d_{\kappa}  \frac{\omega}{D}+2 \frac{D}{\kappa} \frac{\omega^2}{D^2}}{-1-\ii \frac{\omega}{D}+2 \frac{D}{\kappa} \frac{\omega^2}{D^2}},\label{gainf} \\   D&=\frac{1}{\sqrt{G}+1} \frac{\kappa_{\rm id} \kappa_{\rm si}}{\kappa_{\rm id}+\kappa_{\rm si}},
\end{align}
\end{subequations}
where the total coupling rate $\kappa=\kappa_{\rm si}+\kappa_{\rm id}$, the coupling asymmetry $d_{\kappa}=(\kappa_{\rm id}-\kappa_{\rm si})/\kappa$ and the power gain $\sqrt{G}=|g(0)|=(\kappa_{\rm si}\kappa_{\rm id}+4|\lambda|^2)/(\kappa_{\rm si}\kappa_{\rm id}-4|\lambda|^2).$

In the limit of a wide amplification bandwidth $D$, \textit{i.e.}, when the input signals are slowly changing with respect to the time scale $D^{-1}$,  the gain factor $g(\omega)$~\eqref{gainf} is a frequency independent constant to leading order (the zeroth order) in $\omega/D$. This implies time-local input-output relations:
\begin{subequations}\label{ainout}
\begin{align}
\hat{a}_{\rm si, out}(t)&=\sqrt{G}\hat{a}_{\rm si, in}(t)+\sqrt{G-1}\hat{a}^\dagger_{\rm id, in}(t),\\
\hat{a}_{\rm id, out}(t)&=\sqrt{G}\hat{a}_{\rm id, in}(t)+\sqrt{G-1}\hat{a}^\dagger_{\rm si, in}(t), 
\end{align}\end{subequations}expressed in the frame rotating at the resonance frequencies; see Fig.~\ref{JPC_to_BS}(a). For simplicity we have ignored the phase between the idler and the signal ports. The next-to-leading order contribution is to ignore the second order terms $\omega^2/D^2$ but keep the first order terms in Eq.~\eqref{gainf}, resulting in temporally non-local but causal responses with a delay kernel of the type $\int_0^\infty\ee^{-D \tau} \oa_{\rm si/id, in}(t-\tau)\D{\tau}$ in Eqs.~\eqref{ainout}.

\subsection{Erasure of the which-path information}
We see from the input-output relations~\eqref{ainout}, that when the idler input is the vacuum,  both quadratures of the signal input are amplified along with an added extra half a quantum of noise originating from the idler. When both input ports contain signals, they are coherently superposed in the outputs and the which-path information of the signals is erased in the frequency domain.

To further analyze the erasure of the which-path information, we specify that the signal output is measured by heterodyne detection~\cite{WisemanQ, Steck}: equal sampling of quadratures $I \propto \textrm{Re}\expes{\oa_{\rm si, out}}$ and $Q\propto \textrm{Im}\expes{\oa_{\rm si, out}}$, see Fig.~\ref{JPC_to_BS}(a). For simplicity we scale the signal with $1/\sqrt{G}$ and consider the high gain limit $G\gg 1$. The outcomes of a weak continuous measurement of infinitesimal duration $\D{t}$ are,
\begin{subequations}
\label{IQ_pure}
  \begin{align}
\D{I}_{\rm r}(t)&=\sqrt{2}\textrm{Re}\expe{\hat{a}_{\rm si,in}+\hat{a}_{\rm id,in}}\D{t}+\D{W}_{\rm I}(t)\text,\\
\D{Q}_{\rm r}(t)&=\sqrt{2}\textrm{Im}\expe{\hat{a}_{\rm si,in} -\hat{a}_{\rm id,in}}\D{t}+\D{W}_{\rm Q}(t)\text.
\end{align}
\end{subequations}
They consist of two parts: the part expected based on the prior knowledge of the system and the unexpected part (the `innovation') $\D{W}_{\rm I,Q}$. Alternatively, $\D{W}_{\rm I,Q}$ represent the quantum noise of the channels and are modeled by independent Wiener processes with variance $\D{t}$~\cite{WisemanQ, Steck}.

From Eq.~\eqref{IQ_pure}, we see that the amplification-detection scheme is equivalent to a $50$-$50$ beam-splitter followed by phase-sensitive amplifiers in both of the output arms implementing two single-quadrature (homodyne) measurements in the $I$ and $Q$ directions, see Fig.~\ref{JPC_to_BS}. This interpretation also illustrates the nature of which-path erasure: observers of the output ports cannot know where the signals came from. In the high gain limit, the idler output is fully entangled with the signal output~\cite{Flurin} containing no extra information. This can be understood as two-mode squeezing by the amplifier having effectively erased two of the four incoming quadratures. 

The input-output relations~\eqref{ainout} of the amplifier are expressed with the Hermitian conjugated input operators, $\hat{a}^\dagger_{\rm id, in}$ and $\hat{a}^\dagger_{\rm si, in}$. But notice that they can be interpreted as complex conjugation in~Eq.~\eqref{IQ_pure} when the amplifier is followed by the quadrature measurements. This is consistent with the physical picture of unidirectional information flow from the signal sources. The observer of the unidirectionally traveling signals can only make measurements whose measurement backaction to the system is expressed by the operators $\hat{a}_{j}$, that is, \textit{e.g.}, observations of discrete photon emissions or continuous leaking of the cavity field. The interpretation of the amplification-detection stage through beam-splitters and quadrature measurements gives practical means to handle components of cascaded quantum network with Bogoliubov transformations in their input-output relations~\cite{CarmichaelGardiner93, GoughJames09, Gough10}. 

\section{Concurrent generation of remote entanglement} \label{sec:cqed-gen}
\begin{figure}
\centering
\includegraphics[width=1.0\linewidth]{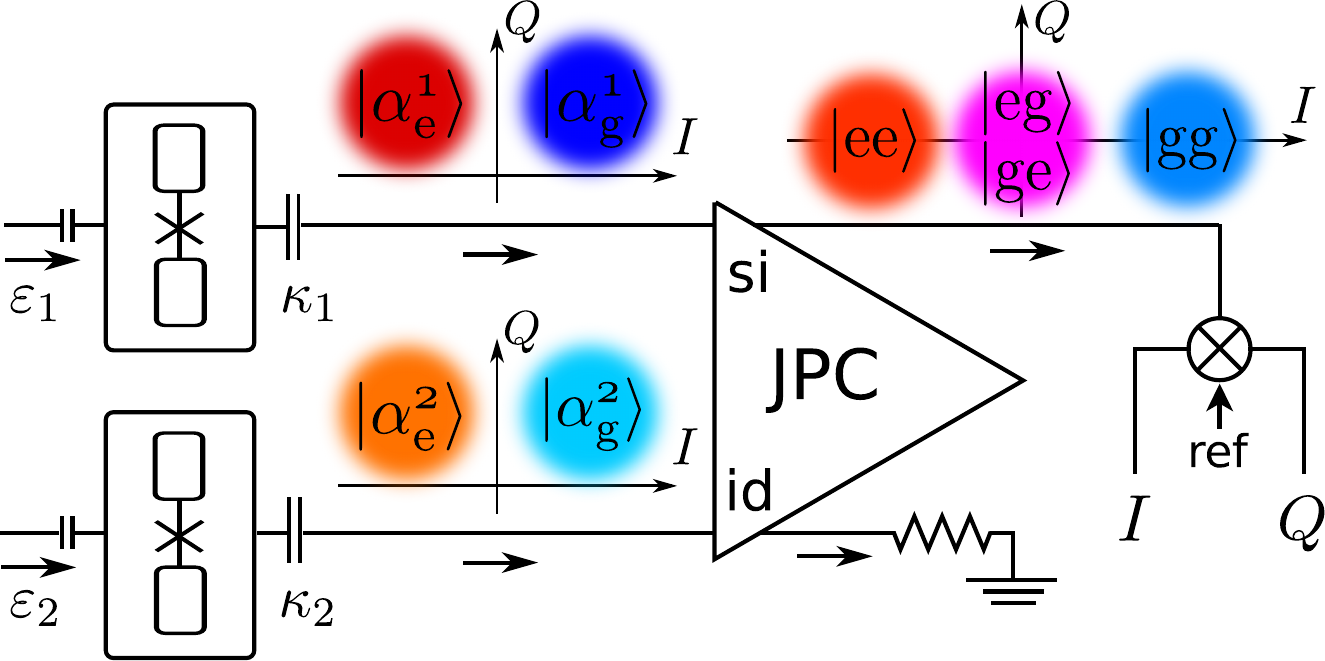}
\caption{\label{cQED_schema}Schematic of a concurrent remote entangling experiment showing two dispersively coupled qubits inside separate, driven and monitored cavities. The left $I$-$Q$ planes show the qubit state dependent coherent states.  The right $I$-$Q$ plane shows the structure of the signal output,~Eq.~\eqref{ainout}, where no which-qubit information is available leading to the joint measurement~$\hat{\sigma}^1_z+\hat{\sigma}^2_z$.}
\end{figure}

We now study a phase-preserving amplifier as a which-path information eraser to concurrently generate remote entanglement.
The considered configuration consists of two transmon qubits inside separate, remote superconducting cavities, see Fig.~\ref{cQED_schema}. The cavities are driven through weakly-coupled input ports and monitored through separate strongly-coupled transmission lines that form the signal and idler ports of a Josephson parametric converter amplifier. In the frame rotating at the cavity driving $\omega^j_{\rm d}=\omega_j-\Delta_j$ and the qubit frequencies $\omega_{\rm q}^j$, the Hamiltonian for a dispersively coupled qubit-cavity pair is 
\begin{equation}
  \hat{H}_j(t)/\hbar= \left(\Delta_{j}+\frac{\chi_{j}}{2}\hat{\sigma}^j_{\rm z}\right)\hat{a}^\dagger_{j} \oa_{j}+\varepsilon_{j}(t)\oa^\dagger_j+\varepsilon^\ast_{j}(t)\oa_j, \label{eq:Hj}
\end{equation}
where $\chi_{j}$ denotes the dispersive coupling strength. When the cavities are driven at their resonance frequencies $\Delta_j=0$, they build up symmetric qubit-state dependent phase responses. In the ideal case these responses are identical and when they are amplified with a high-gain quantum-limited phase-preserving amplifier according to Eq.~\eqref{ainout} there is no which-qubit information left in the outgoing modes. This allows pure joint measurements of the signal sources, here the transmon qubits, with measurement back-action that projects to an entangled subspace.

\subsection{Stochastic master equation of the joint measurement process}
To analyze the configuration and the entanglement generation in detail, it is modeled with a stochastic master equation~\cite{WisemanQ, Steck} (SME) for the density matrix $\orh$ of both qubit-cavity pairs. We derive it by using input-output theory for cascaded quantum systems~\cite{CarmichaelGardiner93, GoughJames09, GardinerCollet85} (essentially $\hat{a}_{\rm si, in}=\sqrt{\kappa_1}\hat{a}_1$ and $\hat{a}_{\rm id, in}=\sqrt{\kappa_2}\hat{a}_2$) and representing the amplification-detection stage as a beam-splitter followed by two quadrature measurements as in Eq.~\eqref{IQ_pure}. See details of the derivation in Appendix~\ref{app:SLH}. The resulting It\={o} stochastic master equation for the monitored qubit-cavity pairs becomes 
\begin{align}
   \D{\orh}=&\sum^2_{j=1}\left\{ \frac{1}{\ii\hbar}[\hat{H}_j(t),\orh]+\mathcal{D}\left(\sqrt{\kappa_{j}}\hat{a}_j\right)\orh \right\}\D{t}\notag \\
&+\sum_{j=1}^2 \left\{\mathcal{D}\left(\sqrt{\Gamma^j}\hat{\sigma}_-^j\right)\orh+\frac 1 2 \mathcal{D}\left(\sqrt{\Gamma^j_\phi}\hat{\sigma}_z^j\right)\orh \right\}\D{t} \notag\\
&+\frac{1}{\sqrt{2}}\mathcal{H}\left( \sqrt{\eta_1\kappa_{1}}\hat{a}_1+\sqrt{\eta_2\kappa_{2}}\hat{a}_2\right)\orh\D{W}_{\rm I}\notag \\
&+\frac{1}{\sqrt{2}}\mathcal{H}\left(\ii\sqrt{\eta_1\kappa_{1}}\hat{a}_1-\ii\sqrt{\eta_2\kappa_{2}}\hat{a}_2\right) \orh\D{W}_{\rm Q}\text,\label{aaSME}
\end{align}
where $\D{\orh}=\orh(t+\D{t})-\orh(t)$. The first two rows describe the open quantum system dynamics. The dissipator terms $\mathcal{D}\left( \hat c \right)\orh= \hat c \orh \hat{c}^\dagger-\frac12 \left \{ \hat{c}^\dagger\hat{c}, \orh \right\}$ model the coupling of the cavity fields $\hat{a}_j$ to the transmission lines with loss rates $\kappa_{j}$. The dissipators for the qubits describe relaxation and pure dephasing with the rates $\Gamma^j=1/T_1^j$ and $\Gamma_\phi^j$. 

The last two rows describe the measurement backaction, which updates the best estimate of the quantum state based on the new information $\D{W}_{{\rm I,Q}}$ in the heterodyne measurement of the signal output. It is represented by innovation terms~\cite{Steck} $\mathcal{H}\left (\hat{c} \right )\orh=\left [\hat{c}-\Tr{\hat \rho \hat c }\right ]\orh+\orh \left [\hat{c}^\dagger-\Tr{\hat \rho \hat{c}^\dagger} \right]$ that are linear in the measurement operators $\hat c$ but nonlinear in the density matrix $\orh$. The efficiencies $\eta_j$, appearing in the innovation terms, describe the fraction of the information measured by the observer. The remaining fraction of the information is lost to the environments and averaging over it leads to dephasing. Here the efficiencies are $\eta_1=\eta\bar{\eta}_{1}$ and $\eta_2=\eta_{\rm g}\eta\bar{\eta}_{2}$. They consist of the transmission coefficients $\bar{\eta}_j$ and the measurement efficiency $\eta$ of the amplification-readout chain. With a finite amplification gain $G$, there is an asymmetry between the idler and signal outputs in Eq.~\eqref{ainout}. When measuring only the signal output, the associated loss of information is represented with the inefficiency $1-\eta_{\rm g}=1/G$ (in practice, $G \sim 10-20$ dB~\cite{Hatridge13, Abdo11}).

\subsection{Stochastic master equation for~the~monitored qubits}
Given that we consider  driven and damped dispersively coupled qubit-cavity systems, the cavity states can be assumed as a superpositions of the qubit dependent coherent states~\cite{Gambetta08} $\{\ket{\alpha^j_{{\rm g}}}\ket{{\rm g}_j},\ket{\alpha^j_{{\rm e}}}\ket{{\rm e}_j}\}$, see Fig.~\ref{cQED_schema}. The coordinates follow the classical equations of motion, $\dot{\alpha}^j_{\rm e,\rm g}(t)=-\ii\varepsilon_{j}(t)-\ii\left(\Delta_j\pm\frac{\chi_j}{2}-\ii\frac{\kappa_{j}}{2}\right)\alpha^j_{\rm e,\rm g}(t)$, see Fig.~\ref{fig3}(c)-(d). With modern superconducting technology, one can achieve typical values of $T_1\sim 100$~$\mu$s~\cite{Paik11, Rigetti12} for the decay times of the qubit. In this regime, the probability for a $T_1$ relaxation event in either of the qubits can be assumed negligible small during the measurement time $T_{\rm m}\sim \kappa^{-1}\sim 1$~$\mu$s. Then the qubits' dynamics can be seen as frozen out, except for the measurement backaction. In this case, the cavity dynamics can be integrated out from the full SME~\eqref{aaSME} by using these time-dependent coherent states as an Ansatz~\cite{Gambetta06, Gambetta08}. Reduction of the full SME~\eqref{aaSME} is very helpful since we are primarily interested in the qubits and the full SME is rather inconvenient to deal with due to its large Hilbert space.

We approach here the integration of the cavity dynamics from a simple point of view. First, by exploiting the linearity of the innovation terms with respect to the measurement operators, we notice that the qubit-cavity pairs in Eq.~\eqref{aaSME} can be organized so that the two blocks look superficially decorrelated, except for the same stochastic `driving fields' $\D{W}_{\rm I, Q}$. In practice, Eq.~\eqref{aaSME} encapsulates a technical source of correlation through the terms $\expes{\oa_j}\orh\D{W}_{\rm I,Q}$ and $\expes{\oa^\dagger_j}\orh\D{W}_{\rm I,Q}$ of the innovation operators. However, as these terms stem from the normalization condition, we can first uncorrelate the two blocks by sacrificing the normalization.  Then, we make use of the single qubit-cavity results of Ref.~\cite{Gambetta08} for integrating out the cavity individually on both qubit-cavity pairs, and finally add the normalization. We have verified these heuristic arguments with a numerical comparison and with an analytic calculation where we explicitly integrate out the cavity dynamics from the full SME~\eqref{aaSME} using a positive P representation~\cite{Gambetta06, WallsMilburn}.

The resulting SME for the qubits' density matrix $\orh_{\rm q}=\textrm{Tr}_{\rm c_{1,2}}\orh$ is
\begin{align}
   \D{\orh_{\rm q}}=&\sum_{j=1}^2\left\{ \frac1{2\ii}\left[\Omega_j(t)\hat{\sigma}^{j}_z,\orh_{\rm q}\right] + \frac 1 2 \mathcal{D}\left(\sqrt{\Gamma^j_{\rm d}(t)}\hat{\sigma}^j_z\right)\hat{\rho}_{\rm q}\right \}\D{t}\notag \\
&+\sum_{j=1}^2 \left\{\mathcal{D}\left(\sqrt{\Gamma^j}\hat{\sigma}_-^j\right)\orh_{\rm q} + \frac 1 2 \mathcal{D}\left(\sqrt{\Gamma^j_\phi}\hat{\sigma}^j_z\right)\hat{\rho}_{\rm q}\right \}\D{t}\notag \\
&+\frac{1}{2\sqrt{2}}\mathcal{H}\left(\mathcal{S}_1(t)\hat{\sigma}^1_z+\mathcal{S}_2(t)\hat{\sigma}^2_z\right)\orh_{\rm q} \D{W_{\rm I}} \notag  \\&+\frac{1}{2\sqrt{2}}\mathcal{H}\left(\ii\mathcal{S}_1(t)\hat{\sigma}^1_z-\ii \mathcal{S}_2(t)\hat{\sigma}^2_z\right)\orh_{\rm q} \D{W_{\rm Q}}\text, \label{qqSME}
\end{align}
where $\Omega_j(t)=\chi_j \textrm{Re}\left\{\alpha^j_{\rm g}(t)[\alpha^j_{\rm e}(t)]^\star\right\}$ is the ac~Stark effect induced by the photons in the cavities. The photons leaking out from the cavity carry information about the qubits' population~\cite{Gambetta06} causing measurement induced dephasing at the~rates~$\Gamma^j_{\rm d}(t)=\chi_j \textrm{Im} \left\{\alpha^j_{\rm g}(t)[\alpha^j_{\rm e}(t)]^\star \right\}$. The information is encoded into the distinguishability of the pointer states $\ket{\alpha^{j}_{{\rm e,g}}}$, see Fig.~\ref{cQED_schema}-\ref{fig3}. Therefore we define the complex measurement amplitude $\mathcal{S}_j(t)$ of the operator $\hat{\sigma}_{z}^j$ as,
\begin{equation}
\mathcal{S}_j(t)\equiv \sqrt{\kappa_{j}\eta_{\rm j}}[\alpha^j_{\rm e}(t)-\alpha^j_{\rm g}(t)]\text, \label{msmt-rate}
\end{equation}
whose real and imaginary parts are related to the measurements in the $I$ and $Q$ directions, respectively. The measurement rate is $\Gamma^j_{\rm m}(t)=|\mathcal{S}_j(t)|^2$.  

The outcomes of a weak continuous measurement of infinitesimal duration of both $I$ and $Q$ quadratures (heterodyne) of the amplified signal, from Eq.~\eqref{IQ_pure}, can be expressed as
\begin{subequations}
\label{eq:qqMSMT}
\begin{align}
  \D{I}_{\rm r}(t)=&\frac{1}{\sqrt{2}}\expe{\textrm{Re}\, \mathcal{S}_1(t)\hat{\sigma}^1_z+\textrm{Re}\,  \mathcal{S}_2(t)\hat{\sigma}^2_z}\D{t}+\D{W}_{\rm I}\notag\\
&+\frac{1}{\sqrt{2}}\textrm{Re}\, \mathcal{U}_1(t)\D{t}+\frac{1}{\sqrt{2}}\textrm{Re}\, \mathcal{U}_2(t)\D{t},\\
  \D{Q}_{\rm r}(t)=&\frac{1}{\sqrt{2}}\expe{\textrm{Im}\, \mathcal{S}_1(t)\hat{\sigma}^1_z-\textrm{Im}\, \mathcal{S}_2(t)\hat{\sigma}^2_z}\D{t}+\D{W}_{\rm Q}\notag\\
&+\frac{1}{\sqrt{2}}\textrm{Im}\, \mathcal{U}_1(t)\D{t}-\frac{1}{\sqrt{2}}\textrm{Im}\, \mathcal{U}_2(t)\D{t},
\end{align}
\end{subequations}
where the expectation values, for example $\expe{\textrm{Re}\, \mathcal{S}_1(t)\hat{\sigma}^1_z}$, are taken for the instantaneous qubits' density matrix $\orh_{\rm q}(t)$. Additionally, the terms involving $\mathcal U_j(t)=\sqrt{\eta_j \kappa_j}[\alpha^j_{\rm e}(t)+\alpha^j_{\rm g}(t)]$ are not informationally meaningful since they only deterministically offset the signals, and will be ignored in what follows. Importantly, SME~\eqref{qqSME} and the measurement outcomes~\eqref{eq:qqMSMT} show that heterodyne measurement of the output of a quantum-limited phase-preserving amplifier implements a pair of two-qubit measurements corresponding to the operators 
\begin{subequations}
  \begin{align}
    \hat{A}_{\rm I}(t)&=\textrm{Re}\,\mathcal{S}_1(t)\hat{\sigma}^1_z+\textrm{Re}\,\mathcal{S}_2(t)\hat{\sigma}^2_z,\\
    \hat{A}_{\rm Q}(t)&=\textrm{Im}\,\mathcal{S}_1(t)\hat{\sigma}^1_z-\textrm{Im}\,\mathcal{S}_2(t)\hat{\sigma}^2_z.
  \end{align}
\end{subequations}
The complex phase and the magnitude of the measurement amplitudes $\mathcal S_j(t)$ are tunable \emph{in situ} by the cavity driving. Thus, the measurement operators can be changed continuously from a simultaneous separate readout of $\hat{A}_{\rm I} \propto \hat{\sigma}_{z}^1$ and $\hat{A}_{\rm Q} \propto \hat{\sigma}_{z}^2$ into a joint entangling readout $\hat{A}_{\rm I} \propto \hat{\sigma}^1_z\pm\hat{\sigma}^2_z$.

\subsection{Balanced driving for perfect erasure of the which-path information from dissimilar sources}
\begin{figure}
\includegraphics[width=1.0\linewidth]{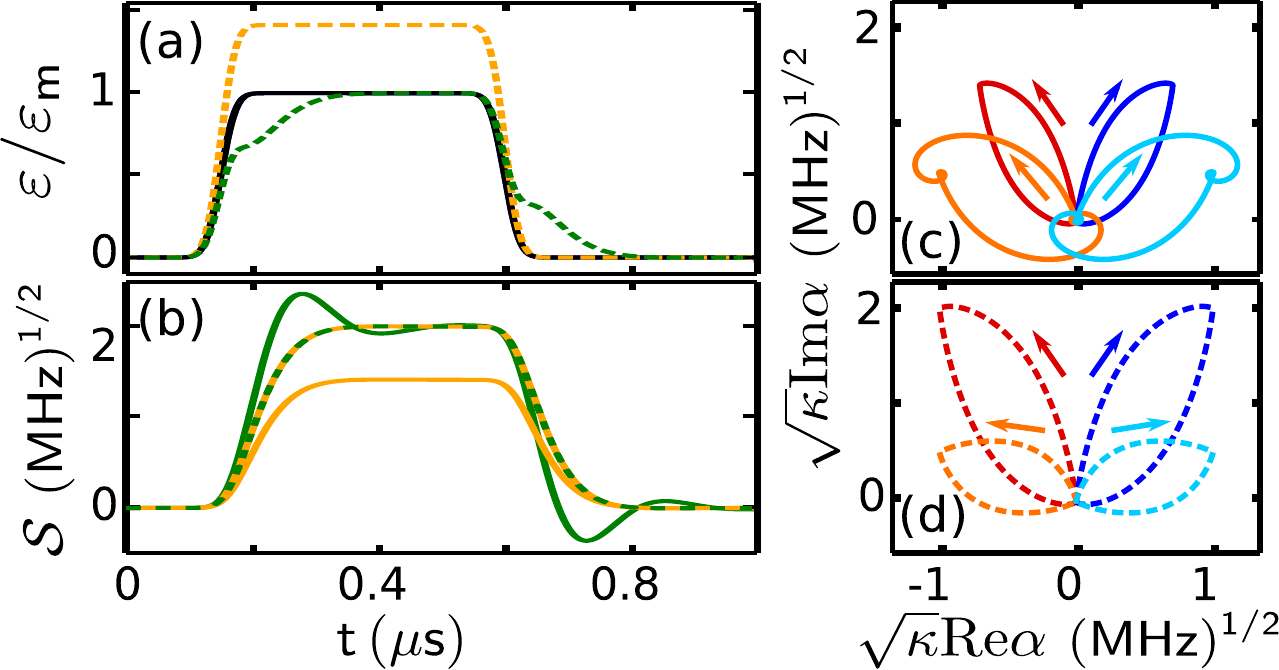}
\caption{\label{fig3} Dynamics of two dissimilar cavities in the frame rotating at the cavity frequencies; (a) Driving amplitudes: unbalanced $\varepsilon_1=\varepsilon_2$~(black), balanced $\varepsilon_{\rm 1}^{(\rm bal)}$~(yellow) and $\varepsilon_{\rm 2}^{(\rm bal)}$~(green); (b) Measurement amplitudes $\mathcal{S}_1$~(yellow) and $\mathcal{S}_2$~(green) where dashed lines refer to driving by the balanced amplitudes; Cavity dynamics for the unbalanced~(c) and balanced~(d) driving amplitudes:  $\sqrt{\kappa_1}\alpha^1_{\rm e}$~(red), $\sqrt{\kappa_1}\alpha^1_{\rm g}$~(blue), $\sqrt{\kappa_2}\alpha^2_{\rm g}$~(light blue) and $\sqrt{\kappa_2}\alpha_{\rm e}^2$~(orange). The parameters are $\kappa_1/2\pi=8$~MHz, $\varepsilon_{\rm m}/2\pi=1$~MHz, and $\kappa_1/\chi_1=\chi_2/\kappa_2=\kappa_1/\kappa_2=2$. The dynamics in the $Q$ direction ($\textrm{Im}\, \alpha$) are left unmatched as informationally insignificant.}
\end{figure}

To utilize the readout for remote heralded entangling, we drive the cavities at resonance $\Delta_{j}=0$ implying $\textrm{Im}\, \mathcal{S}_j(t)=0$. For the most efficient entangling readout, we would like make both measurement amplitudes equal $\mathcal{S}(t)=\mathcal{S}_1(t)=\mathcal{S}_2(t)$ throughout the measurement\----including cavity transients and unequal cavity-qubit parameters. In an entangling readout, one does not want to gather any single-qubit information. Remarkably, the matching of the measurement amplitudes can be achieved by simple engineering of the drive amplitudes. This result  improves the flexibility and scalability of the concurrent remote entangling scheme. We get the balanced driving amplitude,
\begin{equation}\label{eq:balanced}
\varepsilon^{(\rm bal)}_{\rm 2}(t)=\sqrt{\frac{\chi^2_2\eta_{\rm 2}\kappa_{2}}{4}}\left[\kappa_{\rm 2}\dot{\mathcal{S}}_1(t)-\ddot{\mathcal{S}}_1(t)-\frac{\kappa_{2}^2+\chi^2_{2}}4\mathcal{S}_1(t)\right]
\end{equation}
as a result of solving the input of the cavity 2 for a given output $\mathcal{S}_2(t)=\mathcal{S}_1(t)$, visualized in Fig.~\ref{fig3}. Cavity responses in the $Q$ direction are left unmatched because they carry no information about the qubits' population, see Figs.~\ref{cQED_schema}-\ref{fig3}. However, the noise $\D{W}_{\rm Q}$ needs to be recorded because it encodes the stochastic relative phase shift between the qubits due to the unequal photon shot noise in each cavity~\cite{Hatridge13}.  

As the erasure of the which-qubit information in the amplifier output is made perfect  with the balanced driving, the subsequent heterodyne measurement realizes a measurement of the joint operator $\hat{\sigma}_{z}^1+\hat{\sigma}_{z}^2$. For the initial state $\orh_{\rm i}=\ketbra{++}{++}$, where $\hat{\sigma}_{x}\ket{+}=\ket{+}$, the measurement backaction projects the system into a heralded entangled state with the success probability~$p_{\rm s}=1/2$.

\section{The quantum filter} \label{sec:qfilter}
In the quantum filtering, one assumes that the measurement records $\D{I}_r(t)$ and $\D{Q}_r(t)$ are known from the initialization up to a time $t$ and then one would like to know the best estimate of the state of the open quantum system conditioned on this particular measurement record and the initial condition. The stochastic master equation gives the incremental update $\D{\orh_{\rm q}}$ from $\orh_{\rm q}(t)$ to $\orh_{\rm q}(t+\D{t})$ given the new information $\D{W}_{\rm I,Q}(t)$ in the measurement records and knowledge of the system $\orh_{\rm q}(t)$. Naturally, a way to obtain the quantum filter is to solve the SME numerically for each measurement trajectory individually. However, this may generally be a computationally expensive and slow task. Thus, an analytic quantum filter would be much more appealing.

We now consider SME~\eqref{qqSME} with the balanced real measurement amplitudes $\mathcal{S}(t)=\mathcal{S}_1(t)=\mathcal{S}_2(t)$,
\begin{align}
   \D{\orh_{\rm q}}&= \sum_{j=1}^2\frac1{2\ii}\left[\Omega_j(t)\hat{\sigma}^{j}_z,\orh_{\rm q}\right]\D{t}+ \ii \frac{\mathcal{S}(t)}{2\sqrt{2}}[\hat{\sigma}^1_z-\hat{\sigma}^2_z, \orh_{\rm q}]\D{W_{\rm Q}}\notag \\
+&\sum_{j=1}^2 \left\{\mathcal{D}\left(\sqrt{\Gamma^j}\hat{\sigma}_-^j\right)\orh_{\rm q} +  \frac 1 2 \mathcal{D}\left(\sqrt{\Gamma^j_\phi+\Gamma^j_{\rm d}(t)}\hat{\sigma}^j_z\right)\hat{\rho}_{\rm q}\right \}\D{t}\notag \\
+&\frac{\mathcal{S}(t)}{2\sqrt{2}}\mathcal{H}\left(\hat{\sigma}^1_z+\hat{\sigma}^2_z\right)\orh_{\rm q} \D{W_{\rm I}} \text, \label{qqSME_bal}
\end{align}
where we have explicitly written the backaction of the $Q$-measurement in the form of stochastic phase rotation (the second term). The measurement currents are 
\begin{subequations}
\label{eq:msmt-outcomes} 
  \begin{align}
    \D{I}_{\rm r}(t)&=\frac{\mathcal{S}(t)}{\sqrt{2}}\expe{\hat{\sigma}^1_z +\hat{\sigma}^2_z}\D{t}+\D{W}_{\rm I},\\
    \D{Q}_{\rm r}(t)&=\D{W}_{\rm Q}.    
  \end{align}
\end{subequations}
In the following we take into account only the increased dephasing rate of the qubits by the $T_1$ relaxation processes, $\Gamma_2^j=\frac{\Gamma^j}{2}+\Gamma_\phi^j$, but ignore its effect on evolution of the qubits' population. This is justified by long typical $T_1$ times, $T_1\sim 100\ \mu$s, with respect to typical measurement time $T_{\rm m}\sim 1\ \mu$s. 

To derive the quantum filter, one needs to apply the It\={o} rule for changing variables in stochastic calculus~\cite{WisemanQ, Steck}; If $\D{X}=\nu(t) \D{t}+\sigma(t)\D{W}$ and $F(X,t)$, then
\begin{equation}
  \D{F}=\left(\frac{\partial F}{\partial t}+ \nu(t) \frac{\partial F}{\partial X} + \frac{\sigma^2(t)}{2}\frac{\partial^2 F}{\partial X^2}\right)\D{t}+\sigma(t) \frac{\partial F}{\partial X} \D{W}.
\end{equation}
Based on this It\={o} calculus, we have found the analytic solution for SME~\eqref{qqSME} that expresses the two-qubit state $\orh_{\rm q}(t)$ conditioned on an actual stochastic measurement record $[I_{\rm m}(t), Q_{\rm m}(t)]$ and the initial state $\orh_{\rm i}=\ket{++}\bra{++}$. The time-dependent full solutions for the most important two-qubit Bloch coordinates are (the rest are shown in Appendix~\ref{app:bloch})
\begin{subequations} \label{bloch-solutions}
  \begin{align}
    \expe{ZZ}(t)=&\frac{\ee^{-\Lambda(t)} \cosh I_{\rm m}(t)-1}{\ee^{-\Lambda(t)} \cosh I_{\rm m}(t)+1},\label{ZZ-sol}\\
    \expe{ZI}(t)=&\expe{IZ}(t)=\frac{\ee^{-\Lambda(t)} \sinh I_{\rm m}(t)}{\ee^{-\Lambda(t)} \cosh I_{\rm m}(t)+1}, \label{ZI-sol}\\
    \expe{{X}{X}}(t)=&\ee^{-\left(\frac{\Gamma_{\rm m}(t)}{2\eta_{\rm s}\kappa_{\rm s}}+\frac{1-\eta_{\rm t}}{\eta_{\rm t}}\frac{\Lambda(t)}{2}+\sum_{j=1}^2\Gamma^j_2 t\right)}\\ &\times\frac{\ee^{-\Lambda(t)} \cos \Theta_+(t)+\cos\left[Q_{\rm m}(t)-\Theta_-(t)\right]}{\ee^{-\Lambda(t)}\cosh I_{\rm m}(t)+1}\notag,\\
 \expe{{X}{Y}}(t)=&\ee^{-\left(\frac{\Gamma_{\rm m}(t)}{2\eta_{\rm s}\kappa_{\rm s}}+\frac{1-\eta_{\rm t}}{\eta_{\rm t}}\frac{\Lambda(t)}{2}+\sum_{j=1}^2\Gamma^j_2 t\right)}\\ &\times \frac{\ee^{-\Lambda(t)} \sin \Theta_+(t)+\sin\left[Q_{\rm m}(t)-\Theta_-(t)\right]}{\ee^{-\Lambda(t)}\cosh I_{\rm m}(t)+1},\notag\\
    \expe{{X}{I}}(t)=&2\ee^{-\left(\frac{\Gamma_{\rm m}(t)}{2\eta_1\kappa_{1}}+\frac{1-\eta_{\rm 1}}{\eta_{\rm 1}}\frac{\Lambda(t)}{2}+\Gamma^1_2 t\right)}\\&\times\frac{\ee^{-\frac{\Lambda(t)}{2}}\cosh\frac{I_{\rm m}(t)}{2}\cos\left(\frac{Q_{\rm m}(t)}{2}-\Theta_1(t)\right) }{\ee^{-\Lambda(t)} \cosh I_{\rm m}(t)+1}.\notag
  \end{align}
\end{subequations}
The ac Stark effect induced rotation angles are denoted with $\Theta_j(t)= \int_0^t\Omega_j(\tau)\D{\tau}$ and $\Theta_{\pm}(t)=\Theta_1(t)\pm \Theta_2(t)$. We have used the notation of $(\eta_{\rm s}\kappa_{\rm s})^{-1}=\sum_{j=1}^2(\eta_{j}\kappa_{j})^{-1}$ and defined the combined measurement efficiency $\eta_{\rm t}=\eta_1\eta_2/(\eta_1+\eta_2-\eta_1\eta_2)$ that has an important role in the analysis of the entanglement fidelity. 

We define $\Lambda(t)=\int_0^t\Gamma_{\rm m}(\tau) \D{\tau}=\int_0^t |S(\tau)|^2 \D{\tau}$ as the apparent total information content recorded by the observer up to time $t$. The stochastic measurement records $I_{\rm m}(t)$ and $Q_{\rm m}(t)$ are weighted integrals of the measurement outcomes~\eqref{eq:msmt-outcomes},
\begin{equation}
  I_{\rm m}(t)+\ii Q_{\rm m}(t) =\sqrt{2}\int_0^t\mathcal{S}(\tau)\left[\D{I}_{\rm r}(\tau)+\ii \D{Q}_{\rm r}(\tau)\right]\text.  \label{IQ_filtered} 
\end{equation}
The measurement amplitude $\mathcal{S}(t)$ is the correct relative weighting function\----a matched filter\----between the different time instances of the measurement. The analytic solutions of Eqs.~\eqref{bloch-solutions} and~\eqref{Q-filter} have been verified by comparing them to the numerical solution of the stochastic master equation with perfect overlap within the accuracy of the numerical methods. 

\begin{figure}
  \centering
\includegraphics[width=1.0\linewidth]{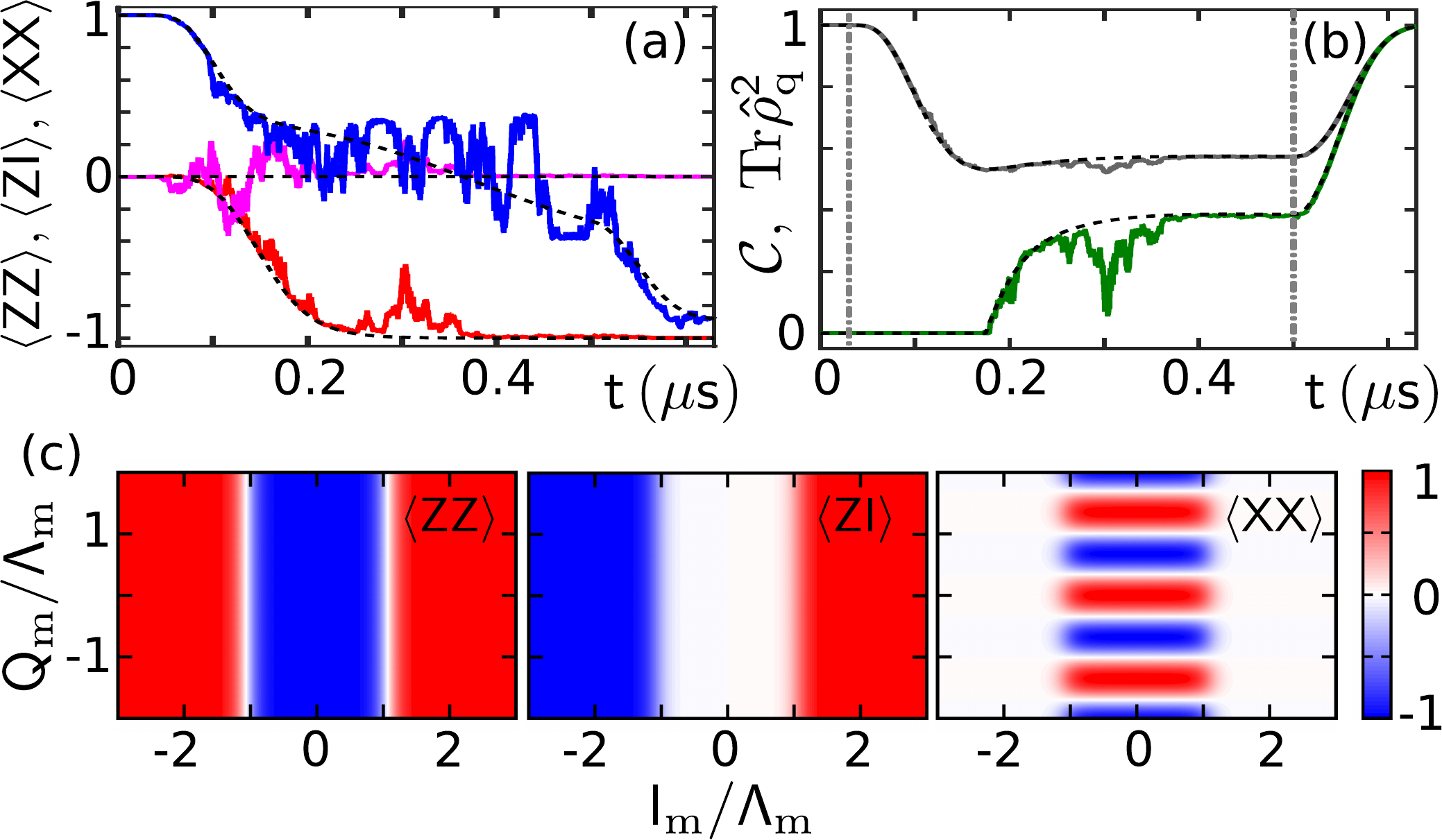} 
\caption{\label{q-trajecs}The panels (a)-(b) show a quantum trajectory given by the quantum filter~\eqref{bloch-solutions} corresponding to a sample measurement record: $\expe{XX}(t)$~(blue), $\expe{ZI}(t)$~(magenta), $\expe{ZZ}(t)$~(red), purity~$\textrm{Tr}\, \orh_{\rm q}^2(t)$~(gray) and concurrence~$\mathcal{C}(t)$~(green). The analytic and numeric (not shown) solutions overlap perfectly. The most probable quantum trajectory~\cite{Jordan13} from the initial $\orh_{\rm i}$ to the final state $\orh_{\rm f}$ is shown as a dashed line; see Appendix~\ref{app:mostprob}. Noise in the $I_{\rm m}$ measurement carries information about the qubits' populations and thus the measurement backaction affects all two-qubit Bloch coordinates. Noise in the $Q_{\rm m}$ measurement carries information only about the relative phase between the qubits, causing fluctuations, \textit{e.g.}, in the $\expes{XX}$ coordinate without affecting to the populations or concurrence. (c) The two-qubit state of Eq.~\eqref{bloch-solutions} shown as as a function of the measurement outcomes~$(I_{\rm m}, Q_{\rm m})$. The parameters are $\kappa_{1}/2\pi=5$~MHz, $\kappa_1/\chi_1=1$, $\kappa_{2}/\kappa_{1}=\chi_{2}/\chi_{1}=1.1$, $\Gamma^j=\Gamma^j_\phi=0$ and $\eta_j=1$. The cavity driving $\varepsilon_{1}(t)$ is on from $t_{\rm s}=0.03$~$\mu$s to $t_{\rm e}=0.5$~$\mu$s [vertical lines in (b)] with the amplitude $\varepsilon_{\rm m}=\kappa_{\rm 1}/\sqrt{6}$ and a temporal shape similar to Fig.~\ref{fig3}(a) resulting in $\Lambda_{\rm m}=\Lambda(T_{\rm m})\approx 3\pi$. The cavity 2 is driven with the balanced driving amplitude ~$\varepsilon^{(\rm bal)}_{2}(t)$ of Eq.~\eqref{eq:balanced}.}
\end{figure}

Quite surprisingly, even for a lossless system, the coherence of a two-qubit state is reduced by the factor $\exp(-\Gamma_{\rm m}(t)/2\eta_{\rm s}\kappa_{\rm s})$~\cite{Kockum12, Wang14, Roch14, Motzoi15}  during the measurement [see Fig.~\ref{q-trajecs}(b)], where
\begin{equation}
 \frac{\Gamma_{\rm m}(t)}{2\eta_{\rm s} \kappa_{\rm s}}=\sum_{j=1}^2 \frac{\Gamma^j_{\rm m}(t)}{2\eta_j \kappa_{j}}=\sum_{j=1}^2\int_0^t\left[\Gamma^j_{\rm d}(\tau)-\frac{\Gamma^j_{\rm m}(\tau)}{2}\right] \D{\tau} \label{non-markov} \text.
\end{equation}
This reduction is due to an accumulated temporal mismatch between the acts that cause the qubit dephasing and the measurement backaction, at the rates $\Gamma^j_{\rm d}(t)$ and $\Gamma^j_{\rm m}(t)/2$, respectively. Physically, each qubit is entangled with cavity photons and the entanglement is not removed by the measurement backaction until the photons have leaked out. At the end of a measurement $t=T_{\rm m}$, when the cavities have been brought back to the vacuum and all the available information has been recorded, the purity revives.  Interestingly, during cavity transients the dephasing rate $\Gamma^j_{\rm d}(t)$ can have negative values, which implies revival of qubit coherence. This non-Markovian dynamics originates from coupling the qubits to Markovian reservoirs indirectly through the cavities. The solutions~\eqref{bloch-solutions}-\eqref{IQ_filtered} generalize and go beyond the previous results~\cite{Gambetta08, Kockum12, Hatridge13, Roch14, Motzoi15, Wang14} by deriving the quantum filter for a concurrent entangling two-qubit readout and verifying the purity reduction directly from stochastic calculus.

The solution~\eqref{bloch-solutions}-\eqref{IQ_filtered} is an analytic quantum filter without need for stochastic numerical solutions that would be limited by time step approximations~\cite{KlodenPlaten, Roch14}. The quantum filter can be interpreted in two ways. First, given an actual measurement record $[I_{\rm m}(t), Q_{\rm m}(t)]$ it draws the stochastic quantum trajectories of the two-qubit state, see Fig.~\ref{q-trajecs}(a)-(b).  In another interpretation, the quantum filter gives the two-qubit state as function of measurement outcomes $(I_{\rm m}, Q_{\rm m})$ representing the effect of the measurement backaction, visualized in Fig.~\ref{q-trajecs}(c). The measurement outcomes $Q_{\rm m}$ are normally distributed with zero mean and variance $\sigma^2=2\Lambda(t)$. For a strong measurement, the distribution of the measurement outcomes $I_{\rm m}$ is a $1:2:1$ mixture of three normal distributions, each with variance $\sigma^2=2\Lambda(t)$, centered at $\bar{I}_{\rm m}=-2\Lambda(t), 0, 2\Lambda(t)$. This gives a definition of the measurement strength $\Delta \bar{I}_{\rm m}/\sigma=\sqrt{2\Lambda(t)}$ as the distinguishability of the parity subspaces.

\subsection{Concurrence}
To examine the fidelity of the entanglement, we calculate the concurrence $\mathcal{C}$~\cite{Wootters98} from the quantum filter solutions [see Fig.~\ref{q-trajecs}(b)]. In the limit of strong measurement $\Delta \bar{I}_{\rm m}/\sigma \gg 1$, the state has collapsed with high probability either to an entangled state with $\expes{ZZ}=-1$ or to a product state with $\expes{ZZ}=1$. In this limit, the concurrence can be accurately approximated from the simplified expression~\cite{ConcurrenceSimp}: $\mathcal{C}(t)=2 \max\left\{0, |\rho_{\rm ge,eg}|-\sqrt{p_{\rm gg,gg}p_{\rm ee,ee}}\right\}$. Expressing this with the quantum filter solutions results in
\begin{equation}
  \mathcal{C}(t) =\max\left\{0,\frac{\exp\left( \frac{3\eta_{\rm t}-1}{\eta_{\rm t}}\frac{\Lambda(t)}{2}-\frac{\Gamma_{\rm m}(t)}{2\eta_{\rm s}\kappa_{\rm s}}-\Gamma^{\rm s}_2 t \right)-1}{\cosh I_{\rm m}(t)+\exp \Lambda(t) }\right\}, \label{eq:concurrence}
\end{equation}
where $\Gamma_2^{\rm s}=\Gamma^1_2+\Gamma^2_2$ is the sum of qubits' dephasing rates. 

Let us now consider the case of where the measurement has ended such that $\Gamma_{\rm m}(t=T_{\rm m})=0$. Curiously, there exists an important bound for the total measurement efficiency $\eta_{\rm t}$ since the numerator of Eq.~\eqref{eq:concurrence} needs to be positive for an entangled state with concurrence $\mathcal{C}>0$. Even for ideal qubits~($\Gamma_2^{\rm t}=0$), there is a threshold $\eta_{\rm t}> 1/3 $ for forming a entangled state. For symmetric transmission this corresponds to $\eta_1=\eta_2 > 1/2$. In the presence of decoherence the bound naturally becomes stricter: 
\begin{equation}
  \eta_{\rm t}> \frac{1}{3-2\frac{\Gamma_2^{\rm s} T_{\rm m}}{\Lambda(T_{\rm m})}.}
\end{equation}
Above the threshold, the purification by the measurement backaction dominates over the measurement induced and qubits' natural dephasing. In addition one can see from Eq.~\eqref{eq:concurrence} that given a fixed measurement time $T_{\rm m}$ and non-ideal efficiencies $\eta_j<1$, there exists an optimal measurement strength $\Lambda_{\rm o}$ that maximizes the concurrence. Physically this can be understood as follows: with the optimal $\Lambda_{\rm o}$, the purification by the measurement backaction is in balance with the measurement induced and qubits' natural dephasing.

\section{Discussion and conclusions} \label{sec:conc}
Compared to corresponding photon-counting based concurrent remote entangling schemes~\cite{Narla16, Kok05} whose entanglement fidelity is more robust to losses and inefficiencies, the proposed continuous variable scheme achieves very high generation rate~$\sim 10^5$ $\text{s}^{-1}$ of entangled qubit pairs ($T_{\rm rep}\sim 5\ \mu$s). Experimental values ($1/\Gamma_2^j=15$~$\mu$s,  $T_{\rm m}=1$~$\mu$s, $\eta_j=0.7$) reachable in near-future superconducting circuit experiments result in concurrence $\mathcal{C} \sim 20 \%$. The current experimental capabilities are such that the cavity-qubit asymmetries can be reduced through our pulse engineering scheme to the point that they will not be a limiting factor\----rather it is the overall efficiency of transmission and measurement, where future technical improvements will lead to considerably better concurrencies.

In conclusion, we considered a readout chain of a quantum-limited phase-preserving amplifier followed by heterodyne detection and developed a physically intuitive description compatible with the theory of cascaded quantum systems. Based on a stochastic master equation approach, we theoretically demonstrated that the amplifier can be utilized as an eraser for the which-qubit information, even from dissimilar sources, and an element in a promising protocol for concurrent entanglement generation between remote superconducting qubits. This protocol is feasible with existing technologies and can be expected to demonstrate formation of entangled remote qubits, primitive constituents of quantum communication and distributed quantum computation. 

\begin{acknowledgments}
We are very grateful for R.~T.~Brierley and Shyam Shankar for many useful discussions. We acknowledge support from ARO W911NF-14-1-0011, W911NF-14-1-0563, NSF DMR-1301798 and the Yale Center for Research Computing. 
\end{acknowledgments}

\appendix
\section{Cascaded quantum systems with a quantum-limited phase-preserving~amplifier and heterodyne detection}\label{app:SLH}
To rigorously derive the stochastic master equation of a cascaded quantum system~\cite{CarmichaelGardiner93}, an effective method is to construct the unidirectional quantum network by using the input-output triplets $G=(S, \hat{L}, \hat{H})$ of the network elements~\cite{GoughJames09}. The $G$-triplet contains the scattering matrix $S$ for the input-output ports of the element, the vector $\hat{L}$ that specifies the coupling to input-output ports and the internal Hamiltonian $\hat{H}$. For compiling a network, one needs to know the rules for the cascade (series) $G_2\vartriangleleft G_1$ and concatenation (parallel) $G_1\boxplus G_2$ products. Let us consider two systems $G_{1}=(S_1,\hat{L}_1,\hat{H}_1)$ and $G_{2}=(S_2, \hat{L}_1, \hat{H}_2)$, then the cascade and concatenation products are, respectively,
\begin{subequations}
\begin{align}
 G_2\vartriangleleft G_1&=\begin{pmatrix} S_2S_1, &S_2\hat{L}_1+\hat{L}_2, & \hat H_{\vartriangleleft} \end{pmatrix},\\
  G_1\boxplus G_2&=\begin{pmatrix} \begin{pmatrix} S_1 & 0\\0 & S_2\end{pmatrix}, &  \begin{pmatrix} \hat{L}_1 \\  \hat{L}_2 \end{pmatrix}, & \hat{H}_1+\hat{H}_2 \end{pmatrix}.
\end{align}
\end{subequations}
In the cascade product, the outputs of the system $G_1$ are connected to inputs of the system $G_2$, and the cascaded Hamiltonian has a corresponding driving term $\hat H_{\vartriangleleft}=\hat{H}_1+\hat{H}_2-\frac{\ii\hbar}{2}(\hat{L}_2^\dagger S_2 \hat{L}_1-\hat{L}_1^\dagger S^\dagger_2 \hat{L}_2)$.

\begin{figure}
  \includegraphics[width=1.0\linewidth]{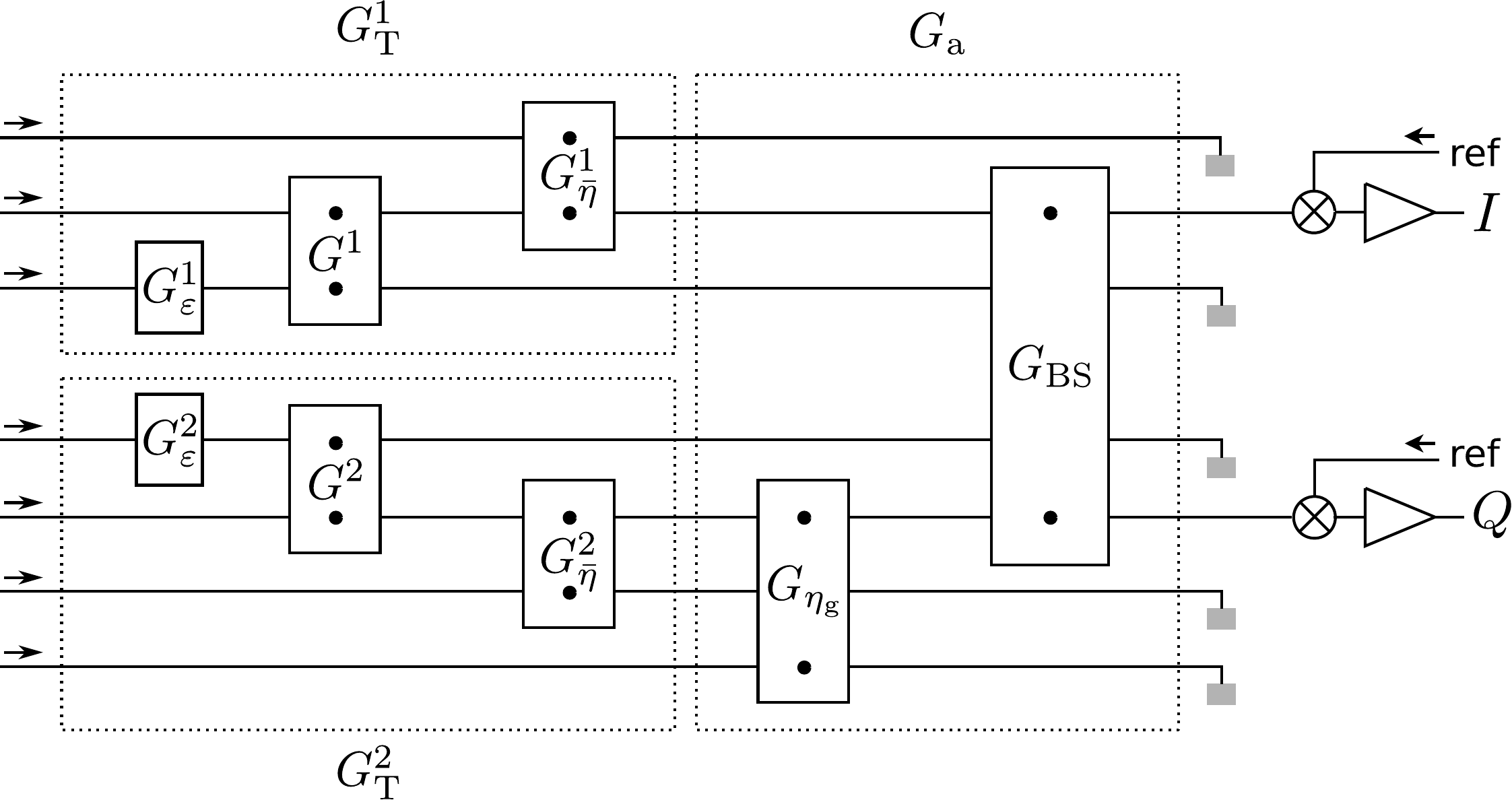}
  \caption{\label{SLH} The concurrent remote entanglement configuration of Fig.~\ref{cQED_schema} in terms of the cascaded triplets $G=\begin{pmatrix}S,&\hat{L},&\hat{H}\end{pmatrix}$. From left to right, the dashed boxes denote the compiled subparts $G_{\rm T}^1$, $G_{\rm T}^2$, and $G_{\rm a}$. The non-monitored channels are denoted with a termination by a gray box. }
\end{figure}

The unidirectional quantum network of the concurrent remote entanglement setup of Figs.\ 1 and 2 is shown in Fig.~\ref{SLH}. The $G$-triplets of the individual elements are:
\begin{subequations}
\begin{align}
  G_{\varepsilon}^j&=\begin{pmatrix}1,& \varepsilon_j(t)/\sqrt{\kappa^{\rm in}_j}, &0 \end{pmatrix}, \\ G^j&=\begin{pmatrix}I_2,& \begin{pmatrix}\sqrt{\kappa^{\rm in}_j}\oa_j \\ \sqrt{\kappa_j}\oa_j\end{pmatrix}, & \hat{H}_j \end{pmatrix}, \\ G^j_{\bar{\eta}}&=\begin{pmatrix}\begin{pmatrix} \sqrt{\bar{\eta}_j},& \sqrt{1-\bar{\eta}_j} \\ -\sqrt{1-\bar{\eta}_j} & \sqrt{\bar{\eta}_j}\end{pmatrix}, &0, &0\end{pmatrix}, \\
G_{\eta_{\rm g}}&=\begin{pmatrix}\begin{pmatrix} \sqrt{\eta_{\rm g}},& 0 &\sqrt{1-\eta_{\rm g}} \\ 0 & 1 & 0 \\-\sqrt{1-\eta_{\rm g}} & 0 & \sqrt{\eta_{\rm g}}\end{pmatrix}, &0, &0 \end{pmatrix},\\
G_{\rm BS}&=\begin{pmatrix}\begin{pmatrix} 1/\sqrt{2}& 0 & 0 & 1/\sqrt{2}  \\ 0 & 1 & 0 & 0\\ 0 & 0 & 1 & 0\\ -1/\sqrt{2} &0 & 0 & 1/\sqrt{2} \end{pmatrix},&  0,& 0  \end{pmatrix},
\end{align}
\end{subequations}
where $G_\varepsilon^j$ denotes the semi-classical driving of the $j$:th cavity, $G^j$ represents the internal dynamics of the $j$:th qubit-cavity pair with the Hamiltonian $\hat{H}_j=\hbar (\Delta_{j}+\frac{\chi_{j}}{2}\hat{\sigma}^j_{\rm z})\hat{a}^\dagger_{j} \oa_{j}$, $G_{\bar{\eta}}^j$ models the losses of the transmission line, $G_{\eta_{\rm g}}$ denotes the effective losses of the amplification-detection stage with $\eta_{\rm g}=(G-1)/G$,  and $G_{\rm BS}$ denotes the effective beam-splitter of the amplification-detection stage shown in Fig.~1. We write the identity (pass-through) element of a single $G_I=(1, 0, 0)$ and several parallel ports $G^{(n)}_I=(I_n, 0, 0)$. The non-monitored channels are denoted with a termination by a gray box in Fig.~\ref{SLH}.

Compilation of the network elements, in accordance with the subparts denoted with dashed boxes in Fig.~\ref{SLH}, results in
\begin{subequations}
\begin{align}
G^1_{\rm T}&=\left(G^1_{\bar{\eta}}\boxplus G_I\right)\vartriangleleft \left(G_I\boxplus G^1\right) \vartriangleleft  \left(G^{(2)}_I\boxplus G^1_\varepsilon\right),\\
  G^2_{\rm T}&= \left(G_I\boxplus G^2_{\bar{\eta}} \boxplus G_I\right) \vartriangleleft \left(G^2 \boxplus G_I^{(2)}\right) \vartriangleleft   \left(G^2_\varepsilon \boxplus G^{(3)}_I\right),\\
G_{\rm a}&= \left(G_I\boxplus G_{\rm BS} \boxplus G_I\right) \vartriangleleft \left(G^{(4)}_I\boxplus G_{\eta_{\rm g}}\right),
\end{align}
\end{subequations}
where $G^j_{\rm T}$ denotes a driven qubit-cavity pair connected into lossy transmission lines and $G_{\rm a}$ is the effective amplification-detection stage by the quantum-limited phase-preserving amplifier and heterodyne detection of one of the output ports.  The full compilation of these subparts gives the total network triplet
\begin{equation}
G^T=G_{\rm a}\vartriangleleft \left(G_{\rm T}^1\boxplus G_{\rm T}^2\right)=\begin{pmatrix}S_T,& \hat{L}_T,& \hat{H}_T\end{pmatrix}.
\end{equation}

To form the stochastic master equation of the full quantum network, we need to know the Hamiltonian $\hat{H}_{\rm T}=\sum_j \hat{H}_j+\hbar\varepsilon_{j}(t)\oa^\dagger_j+\hbar\varepsilon^\ast_{j}(t)\oa_j=\sum_j \hat{H}_j(t)$, which equals to Eq.~\eqref{eq:Hj}, and the coupling vector 
 \begin{align}
   \hat{L}_T=\begin{pmatrix} \sqrt{\kappa_1(1-\bar{\eta}_1)}\oa_1 \\  \frac{1}{\sqrt{2}}\left(\sqrt{\kappa_1\bar{\eta}_1}\oa_1+ \sqrt{\kappa_2\bar{\eta}_2\eta_{\rm g}}\oa_2\right) \\ \sqrt{\kappa^{\rm in}_1}\oa_1 \\ \sqrt{\kappa^{\rm in}_2}\oa_2 \\ \frac{1}{\sqrt{2}}\left( \sqrt{\kappa_1\bar{\eta}_1}\oa_1-\sqrt{\kappa_2\bar{\eta}_2\eta_{\rm g}}\oa_2\right) \\ \sqrt{\kappa_2(1-\bar{\eta}_2)}\oa_2 \\  \sqrt{\kappa_2(1-\eta_{\rm g})}\oa_2  \end{pmatrix},
 \end{align}
which have been simplified by absorbing the driving terms into the Hamiltonians. Finally, the stochastic master equation corresponding to homodyne detection of the channels $2$ and $4$, see Fig.~\ref{SLH}, with the measurement efficiency $\eta$ can be written as
\begin{align} 
 \D{\orh}=\frac{1}{\ii\hbar} &[\hat{H}_{\rm T}, \orh]\D{t}+\sum_{k=1}^7 \mathcal{D}(\hat{L}_k)\orh \D{t}\notag\\&+\mathcal{H}(\sqrt{\eta}\hat{L}_2)\orh\D{W}_{\rm I}+\mathcal{H}(\ii \sqrt{\eta}\hat{L}_4)\orh\D{W}_{\rm Q} \notag \\
=\sum^2_{j=1}&\left\{\frac{1}{\ii\hbar} [\hat{H}_j(t),\orh]+\mathcal{D}\left(\sqrt{\kappa_{j}+\kappa^{\rm in}_j}\hat{a}_j\right)\orh \right\}\D{t}\notag \\ &+\frac{1}{\sqrt{2}}\mathcal{H}\left(\sqrt{\eta_1\kappa_{1}}\hat{a}_1+\sqrt{\eta_2\kappa_{2}}\hat{a}_2\right)\orh\D{W}_{\rm I}\notag\\&+\frac{1}{\sqrt{2}}\mathcal{H}\left(\ii\sqrt{\eta_1\kappa_{1}}\hat{a}_1-\ii\sqrt{\eta_2\kappa_{2}}\hat{a}_2\right)\orh\D{W}_{\rm Q}. 
\end{align}
In the main text, the small difference between $\kappa_j$ and $\kappa_j+\kappa^{\rm in}_{j}$ is omitted for simplicity (or absorbed into the loss coefficients of the transmission lines $\bar{\eta}_j$) and the qubits' relaxation and pure dephasing terms have been added.

\section{The quantum filter}\label{app:bloch} 
Here we list the rest of the two-qubit Bloch coordinates of the analytic quantum filter solution of Eq.~\eqref{bloch-solutions} for SME~\eqref{qqSME_bal}:
\begin{subequations}
\label{Q-filter}
\begin{align}
  \expes{{Y}{X}}(t)=&\ee^{-\left(\frac{\Gamma_{\rm m}(t)}{2\eta_{\rm s}\kappa_{\rm s}}+\frac{1-\eta_{\rm t}}{\eta_{\rm t}}\frac{\Lambda(t)}{2}+\sum_{j=1}^2\Gamma^j_2 t\right)} \\&\times\frac{\ee^{-\Lambda(t)} \sin \Theta_+(t)-\sin\left[Q_{\rm m}(t)-\Theta_-(t)\right]}{\ee^{-\Lambda(t)}\cosh I_{\rm m}(t)+1},\notag \\
  \expes{{Y}{Y}}(t)=&\ee^{-\left(\frac{\Gamma_{\rm m}(t)}{2\eta_{\rm s}\kappa_{\rm s}}+\frac{1-\eta_{\rm t}}{\eta_{\rm t}}\frac{\Lambda(t)}{2}+\sum_{j=1}^2\Gamma^j_2 t\right)} \\&\times\frac{-\ee^{-\Lambda(t)} \cos \Theta_+(t)+\cos\left[Q_{\rm m}(t)-\Theta_-(t)\right]}{\ee^{-\Lambda(t)}\cosh I_{\rm m}(t)+1},\notag \\
  \expes{{X}{Z}}(t)=&2\ee^{-\left(\frac{\Gamma_{\rm m}(t)}{2\eta_1\kappa_{1}}+\frac{1-\eta_{\rm 1}}{\eta_{\rm 1}}\frac{\Lambda(t)}{2}+\Gamma^1_2 t\right)}  \\&\times\frac{\ee^{-\frac{\Lambda(t)}{2}}\sinh\frac{I_{\rm m}(t)}{2}\cos\left(\frac{Q_{\rm m}(t)}{2}-\Theta_1(t)\right) }{\ee^{-\Lambda(t)} \cosh I_{\rm m}(t)+1},\notag \\
 \expes{{Z}{X}}(t)=&2\ee^{-\left(\frac{\Gamma_{\rm m}(t)}{2\eta_2\kappa_{2}}+\frac{1-\eta_{\rm 2}}{\eta_{\rm 2}}\frac{\Lambda(t)}{2}+\Gamma^2_2 t\right)}  \\&\times \frac{\ee^{-\frac{\Lambda(t)}{2}}\sinh\frac{I_{\rm m}(t)}{2}\cos\left(\frac{Q_{\rm m}(t)}{2}+\Theta_2(t)\right)}{\ee^{-\Lambda(t)} \cosh I_{\rm m}(t)+1},\notag \\
  \expes{{Y}{Z}}(t)=&-2\ee^{-\left(\frac{\Gamma_{\rm m}(t)}{2\eta_1\kappa_{1}}+\frac{1-\eta_{\rm 1}}{\eta_{\rm 1}}\frac{\Lambda(t)}{2}+\Gamma^1_2 t\right)}  \\&\times \frac{\ee^{-\frac{\Lambda(t)}{2}}\sinh\frac{I_{\rm m}(t)}{2}\sin\left(\frac{Q_{\rm m}(t)}{2}-\Theta_1(t)\right) }{\ee^{-\Lambda(t)} \cosh I_{\rm m}(t)+1},\notag \\
 \expes{{Z}{Y}}(t)=&2\ee^{-\left(\frac{\Gamma_{\rm m}(t)}{2\eta_2\kappa_{2}}+\frac{1-\eta_{\rm 2}}{\eta_{\rm 2}}\frac{\Lambda(t)}{2}+\Gamma^2_2 t\right)}  \\&\times \frac{\ee^{-\frac{\Lambda(t)}{2}}\sinh\frac{I_{\rm m}(t)}{2}\sin\left(\frac{Q_{\rm m}(t)}{2}+\Theta_2(t)\right)}{\ee^{-\Lambda(t)} \cosh I_{\rm m}(t)+1},\notag \\
  \expes{{Y}{I}}(t)=&-2\ee^{-\left(\frac{\Gamma_{\rm m}(t)}{2\eta_1\kappa_{1}}+\frac{1-\eta_{\rm 1}}{\eta_{\rm 1}}\frac{\Lambda(t)}{2}+\Gamma^2_1 t\right)}  \\&\times \frac{\ee^{-\frac{\Lambda(t)}{2}}\cosh\frac{I_{\rm m}(t)}{2}\sin\left(\frac{Q_{\rm m}(t)}{2}-\Theta_1(t)\right) }{\ee^{-\Lambda(t)} \cosh I_{\rm m}(t)+1},\notag \\
  \expes{{I}{X}}(t)=&2\ee^{-\left(\frac{\Gamma_{\rm m}(t)}{2\eta_2\kappa_{2}}+\frac{1-\eta_{\rm 2}}{\eta_{\rm 2}}\frac{\Lambda(t)}{2}+\Gamma^2_2 t\right)}  \\&\times \frac{\ee^{-\frac{\Lambda(t)}{2}}\cosh\frac{I_{\rm m}(t)}{2}\cos\left(\frac{Q_{\rm m}(t)}{2}+\Theta_2(t)\right) }{\ee^{-\Lambda(t)} \cosh I_{\rm m}(t)+1},\notag \\
  \expes{{I}{Y}}(t)=&2\ee^{-\left(\frac{\Gamma_{\rm m}(t)}{2\eta_2\kappa_{2}}+\frac{1-\eta_{\rm 2}}{\eta_{\rm 2}}\frac{\Lambda(t)}{2}+\Gamma^2_2 t\right)}  \\&\times \frac{\ee^{-\frac{\Lambda(t)}{2}}\cosh\frac{I_{\rm m}(t)}{2}\sin\left(\frac{Q_{\rm m}(t)}{2}+\Theta_2(t)\right) }{\ee^{-\Lambda(t)} \cosh I_{\rm m}(t)+1} \notag.
\end{align}
\end{subequations}
The notation is the same as introduced in Sec.~\ref{sec:cqed-gen}.

\section{The most probable trajectory to an entangled state}\label{app:mostprob}
Given the initial  $\orh_{\rm i}=\ketbra{++}{++}$ and final state $\orh(T_{\rm m})=\orh_{\rm f}$, as well as the time evolution in between described by SME~\eqref{qqSME_bal}, one may ask which are all the measurement trajectories $\{I_{\rm r}, Q_{\rm r} \}$ that connect the boundary conditions $\orh_{\rm i}$ and $\orh_{\rm f}$, and what is the most probable of these trajectories, denoted here $\{\widetilde{I}_{\rm r},\widetilde{Q}_{\rm r}\}$. First, the inversion of the quantum filter of Eqs.~\eqref{bloch-solutions} and ~\eqref{Q-filter} gives the pair of values $(I_{\rm m}, Q_{\rm m})$ corresponding to the final state $\orh_{\rm f}$. Then Eq.~\eqref{IQ_filtered} shows that the initial and final values are connected by all those trajectories $\{ I_{\rm r}, Q_{\rm r}\}$ that produce the pair $(I_{\rm m}, Q_{\rm m})$ via importance weighting~\eqref{IQ_filtered}.  However, this does not answer to the question what is the most probable trajectory. For that purpose, one needs to resort to more advanced methods in general. A possibility is to formulate a probability distribution of the measurement trajectories and apply the SME to make the connection between the time evolution of the quantum state and the measurement trajectories and, finally, by extremizing the distribution one finds the most likely trajectory~\cite{Jordan13}. 

Given the quantum filter of Eqs.~\eqref{bloch-solutions} and~\eqref{Q-filter}, some of the most probable paths can be found by simple probabilistic arguments. Let us concentrate on the measurement in the $I$ direction and consider the trajectories $I_{\rm r}(t)$ that end up in an entangled state for which $\expes{ZZ}=-1$. In the strong measurement limit, the end points of the weighted trajectories $I_{\rm m}(t)$ are normally distributed around three means $-2\Lambda(t)$,  $0$, $2\Lambda(t)$. All the trajectories $I_{\rm m}(t)$, whose endpoint $I_{\rm m}(T_{\rm m})$ is around the origin, correspond to an entangled state $\expes{ZZ}=-1$, see Fig.~\ref{q-trajecs}(c). Then the most probable end point producing an entangled state is $I_{\rm m}(T_{\rm m})=0$. To see what is the most probable of all the trajectories $I_{\rm r}(t)$ that give rise to $I_{\rm m}(T_{\rm m})=0$ we remember that the initial state is $\expes{ZI}(t=0)=0$ and the weighting equation is,
\begin{align}
 \D{I_{\rm m}}(t)&=\sqrt{2}\mathcal{S}(t)\left(\sqrt{2}\mathcal{S}(t)\expes{ZI}(t)\D{t} + \D{W}_{\rm I}(t)\right)\notag \\&=\sqrt{2}S(t)\D{I}_{\rm r}(t).
 \label{dIm}
\end{align}
This shows that, in general, the probability cost of generating an update of the quantum trajectory is minimized by choosing identically $\D{W}_{\rm I}(t)=0$. For this update and the initial state $\ketbra{++}{++}$, SME~\eqref{qqSME_bal} keeps $\expes{ZI}(t)=0$ and gradually updates $\expes{ZZ}$ towards $\expes{ZZ}=-1$. Thus by taking this update $\D{W}_{\rm I}(t)=0$ at every time step, the trajectory is $I_{\rm r}(t)=0$ and it results in the most probable end point $I_{\rm m}(T_{\rm m})=0$. Based on this, the most probable trajectory to the most probable entangled state is identically zero $\widetilde{I}_{\rm r}(t)=0$.

For the measurement in the $Q$ direction, the end point $Q_{\rm m}(T_{\rm m})$ corresponding to the final state is solved again from the quantum filter of Eqs.~\eqref{bloch-solutions} and~\eqref{Q-filter}. Then from the weighting equation,
\begin{equation}
  \D{Q_{\rm m}}(t)=\sqrt{2}\mathcal{S}(t)\D{W}_{\rm Q}(t)=\sqrt{2}\mathcal{S}(t)\D{Q}_{\rm r}(t), \label{dQm}
\end{equation}
we see that the probability cost to reach the final value is minimized by updating the trajectory by $\D{Q_{\rm r}(t)}=Q'_{\rm m}\mathcal{S}(t)\D{t}/\sigma^2$, where $Q'_{\rm m}=Q_{\rm m}(T_{\rm m}) \mod 2\pi$ and $\sigma^2=\sqrt{2}\int_0^{T_{\rm m}} \mathcal{S}^2(\tau)\D{\tau}$.  This gives rise to the most probable trajectory $\widetilde{Q}_{\rm r}(t)=Q'_{\rm m}\int_0^t S(\tau) \D{\tau}/\sigma^2$. In Fig.~\ref{q-trajecs}(a)-(b), we have shown the Bloch coordinates corresponding the most probable measurement trajectory as a dashed lines.

\end{document}